# Solar Magnetic Flux Ropes

Boris Filippov[1], Olesya Martsenyuk[1], Abhishek K. Srivastava[2], and Wahab Uddin[3]

[1]*Pushkov Institute of Terrestrial Magnetism, Ionosphere and Radio Wave Propagation, Russian Academy of Sciences (IZMIRAN), Troitsk, Moscow, 142190, Russia*

[2]*Department of Physics, Indian Institute of Technology (Banaras Hindu University), Varanasi 221005, India*

[3]*Aryabhatta Research Institute of Observational Sciences (ARIES), Manora Peak, Nainital 263 002, Uttarakhand, India*

**Abstract.** At the beginning of 1990s, it was found that the strongest disturbances of the space-weather were associated with huge ejections of plasma from the solar corona, which took the form of magnetic clouds when moved from the Sun. It is the collisions of the magnetic clouds with the Earth's magnetosphere that lead to strong, sometimes catastrophic changes in space-weather. The onset of a coronal mass ejection (CME) is sudden and no reliable forerunners of CMEs have been found till date. The CME prediction methodologies are less developed compared to the methods developed for the prediction of solar flares. The most probable initial magnetic configuration of a CME is a flux rope consisting of twisted field lines which fill the whole volume of a dark coronal cavity. The flux ropes can be in stable equilibrium in the coronal magnetic field for weeks and even months, but suddenly they loose their stability and erupt with high speed. Their transition to the unstable phase depends on the parameters of the flux rope (i.e., total electric current, twist, mass loading etc.), as well as on the properties of the ambient coronal magnetic field. One of the major governing factors is the vertical gradient of the coronal magnetic field which is estimated as decay index ($n$). Cold dense prominence material can be collected in the lower parts of the helical flux tubes. Filaments are therefore good tracers of the flux ropes in the corona, which become visible long before the beginning of the eruption. The perspectives of the filament eruptions and following CMEs can be estimated by the comparison of observed filament heights with calculated decay index distributions. The present paper reviews the formation of magnetic flux ropes, their stable and unstable phases, eruption conditions, and also discusses their physical implications in the solar corona.

## 1. Introduction

Solar storms influence the human life to an increasing extent (Clark 2006). This is because the modern civilization uses more and more systems (e.g., communication system) and tools of the global scale (e.g., air aviation, satellites), which are influenced not only by the local environment but also by the conditions in the whole near-Earth space. The totality of factors determining the electromagnetic conditions and radiation situation in the vicinity of Earth is now named as "space weather". Space weather variability is determined mostly by sporadic active processes on the Sun, which disturb the interplanetary medium and then through variations of the solar wind cause the geomagnetic storms.

Solar flares were considered for a long time as the main and pivotal sources of the sudden disturbances of interplanetary medium and geophysical conditions beginning from the date of their discovery in 1859 (Carrington 1859; Hodgson 1859). Solar filament eruptions were other conspicuous phenomena in the low and middle corona that were found to have a strong influence on the Earth's outer environment. At the beginning of 1990s with the help of space-borne coronagraphic observations, it was found that the strongest disturbances of space weather were associated with huge ejection of the matter from the solar corona, which took the form of magnetic clouds when moved from the Sun (Kahler 1992; Gosling 1993; Huttunen *et al*. 2002; Gopalswamy 2008). It is the collisions of the magnetic clouds with the Earth's magnetosphere that lead to strong, sometimes to catastrophic, changes in space weather. The total mass involved in motion is up to $10^{16}$ g and the speed is up to 2000 km s$^{-1}$ for exceptionally big CMEs (Gosling *et al*. 1974; Hildner 1977; Jackson & Howard 1993; Vourlidas *et al*. 2002; . Koutchmy *et al*. 2008). Therefore, the energy needed to accelerate this material within about half an hour is approximately $10^{32}$ ergs at most (the corresponding rate of energy transport is then about $10^{29}$ ergs s$^{-1}$). The rate of energy supply is not small. Indeed, it is reaching few percents of the thermal emission of the photosphere, i.e., $6.3 \cdot 10^{10}$ ergs s$^{-1}$ cm$^{-2}$ through an area of $10^4$ Mm$^2$ situated below the erupting volume, namely ~ $10^{31}$ ergs s$^{-1}$. However, this energy cannot be used in the corona because it is transparent for the photospheric radiation. The energy flow needed for accelerating the material, on the other hand, is large compared to the power needed for the coronal and chromospheric heating above this area ~ $10^{27}$ ergs s$^{-1}$ (Withbroe & Noyes 1977). The rate of magnetic energy injection into the corona through the photosphere in an active region was evaluated to ~ $5 \cdot 10^{27}$ ergs s$^{-1}$ (Régnier & Canfield 2006). These facts lead to a concept of energy storage in the corona in the form of free magnetic energy or energy of coronal electric currents.

Coronal electric currents seem to be the only possible form of the storage of energy in the corona that release in eruptive events. Since plasma density is low in the corona, electric currents in the most general case should be field aligned and the magnetic field should be force-free

$$\text{curl } \boldsymbol{B} = \alpha \, \boldsymbol{B} \,. \tag{1}$$

A linear force-free field with a constant α has the minimum energy for given magnetic helicity. Therefore, for the effective storage of free energy, the force-free field should be sufficiently non-linear with α strongly varying in the space. Electric current sheets and magnetic flux ropes are considered as most probable structures that able to store free magnetic energy in the corona (Priest & Forbes 2002; Podgorny & Podgorny 2006; Kliem & Török 2006; Fan & Gibson 2007).

Hyperbolic magnetic configurations around null points of the X-type are believed to be the most suitable place for the current sheets. It seems that the occurrence of the specific hyperbolic magnetic field configuration is the necessary condition for the solar flares and the key is only to locate such configurations in the solar atmosphere. Giovanelli (1947, 1948) was the first who included them into the flare theory. Then, Dungey (1953) pointed out the instability of the magnetoplasma in the vicinity of a null point. Later, Parker (1957), Severny (1958), Sweet (1958, 1969), Petschek (1964), Syrovatsky (1966), Sonnerup (1970), and many others, treated the problem of field annihilation in the X-type magnetic field configuration. For many years theoretical analysis was restricted mainly to 2D geometry for the reasons of simplicity and understanding. Although 2D magnetic reconnection was studied in detail both analytically (Parker 1973; Priest & Forbes 1986; Somov 1986; Jardine & Priest, 1988) and numerically (Biskamp 1982; 1984, Lee and Fu 1986; Scholer 1989), the theory meets with major difficulties in observational verification. 3D reconnection was studied later both analytically and numerically (Priest & Titov 1996; Pontin *et al*. 2004, 2005; Rickard & Titov 1996; Galsgaard & Nordlund 1997; Galsgaard, Priest, & Titov 2003; Pontin & Galsgaard 2007; Pontin, Bhattacharjee, &

Galsgaard 2007; Pariat *et al.* 2009). Observations interpreted as an evidence of current sheets presence in the corona show the appearance of specific structures following onsets of eruptions (Sui & Holman 2003; Lin *et al*. 2005; Liu *et al*. 2009, 2010; Su *et al*, 2013).

Current sheets are considered as very thin structures dividing oppositely directed magnetic fields. In contrast, a flux rope is usually believed to be a volumetric plasma structure with the magnetic field lines wrapping around a central axis. Such configurations were intensively studied in the laboratory plasma physics in connection with nuclear fusion problems. In cylindrical geometry, a straight flux rope is represented by a Z-pinch (Buneman 1961; Zueva, Solov'ev, & Morozov 1976; Lee 1983). In axisymmetric geometry, plasma volume in the shape of a torus is confined by magnetic field lines that move around the torus in a helical shape (Shafranov 1966; Braams & Stott 2002). Magnetic confinement within toroidal plasma tubes was realized in tokamak and stellarator nuclear fusion installations.

## 2. Formation of Magnetic Flux Ropes

Theoretically, magnetic flux ropes can be formed in the corona in two ways: (i) magnetic reconnection of stressed arcades in the corona, and (ii) bodily flux emergence from below the photosphere. In the reconnection model, the imposed boundary movements such as converging and shearing motions of different polarities, rotation of sunspots, and magnetic flux cancellation twist and stretch the initial potential field gradually, leading to magnetic reconnection (Brandt *et al.* 1988; Browning 1991; Amari *et al.* 2011; Aulanier *et al.* 2010). If an electric current is generated in the corona, its magnetic field spreads in all directions until it meets plasma able to resist to the generated magnetic field pressure. In the low density corona, this boundary could be far away from the position of the current. For a typical for solar filaments electric current of $10^{11}$ A (Ballester 1984; Kulikova *et al.* 1986; Vrsnak *et al.* 1988; Srivastava, Ambastha, & Bhatnagar 1991), the boundary is located at a distance of half of a solar radius from the current axis. However the dense photospheric layers do not allow a coronal magnetic field to penetrate into the solar interior. Diamagnetic currents are induced at the surface of the photosphere. These induced photospheric currents are sometimes referred as a mirror current because they produce a magnetic field in the corona equivalent to the magnetic field of the mirror image of the coronal current with an opposite direction (Kuperus & Raadu 1974).

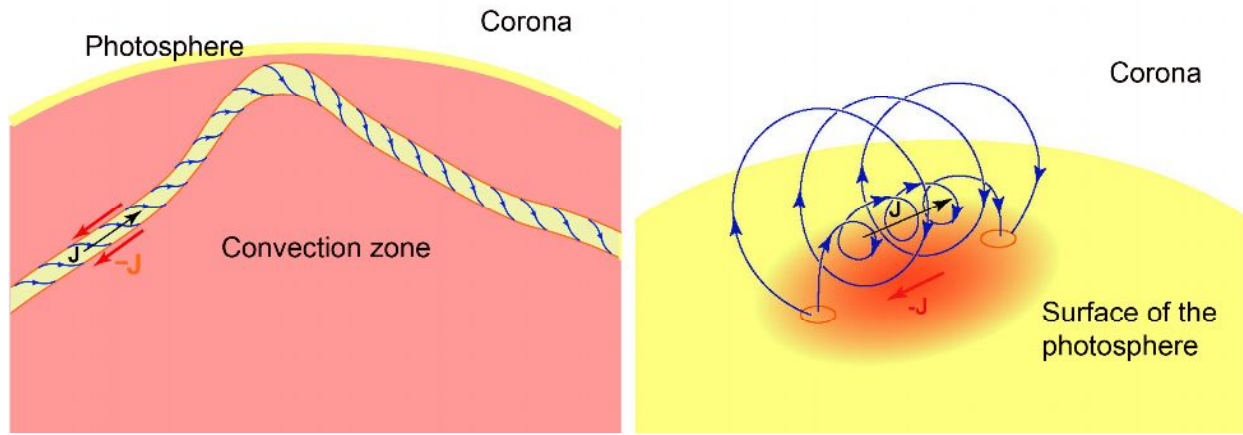

**Figure 1.** Emergence of a twisted flux tube from the convection zone into the corona.

In the emergence model, a twisted flux rope is assumed to exist below the photosphere and emerge into a preexisting coronal potential field (Fan 2001, Fan 2010; Manchester *et al.* 2004; Magara 2006). A twisted flux tube could emerge from the convection zone due to the magnetic buoyancy effect. Below the photosphere, the magnetic field of the flux rope is confined by the shielding oppositely directed boundary current (Parker 1979; Solov'ev 1985), which is held by a plasma pressure gradient (Fig. 1). After rising into the corona, the shielding current shell spreads far in all directions, except the downward direction where it again meets the resistance of the photospheric layers and forms an analog of the mirror current. Longcope & Welsch (2000) suggested, based on a simplified analytical model, that return currents may even completely remain below the corona during the emergence of magnetically isolated flux tubes. Török *et al.* (2014) analyzed the evolution of electric currents during the formation of a bipolar active region by considering a three-dimensional magnetohydrodynamic simulation of the emergence of a sub-photospheric, current-neutralized magnetic flux rope into the solar atmosphere. They found that a strong deviation from current neutralization developed simultaneously with the onset of significant flux emergence into the corona, accompanied by the development of substantial magnetic shear along the active region's polarity inversion line (neutral line). After the region has formed and flux emergence has ceased, the strong magnetic fields in the center of active region are connected solely by direct currents, and the total direct current is several times larger than the total return current. These results suggest that active regions, the main sources of coronal mass ejections and flares, are born with substantial net currents, which is in agreement with recent observations.

Okamoto *et al.* (2008, 2009) reported observations obtained with the Solar Optical Telescope (SOT) on board the *Hinode* satellite indicating that helical flux rope was emerging from below the photosphere into the corona along the polarity inversion line under the preexisting prominence. They suggest that this supply of a helical magnetic flux to the corona is associated with an evolution and maintenance of active region prominences. The observed properties of the developing filament channel studied by Lites *et al.* (2010) are also in accordance with the notion of the emergence through the photosphere of a slightly twisted horizontal flux rope.

## 3. Observational Manifestations of Flux Ropes in the Corona

How can one find observational manifestations of flux ropes in the corona? The coronal magnetic field is still largely elusive for reliable measurements. Photospheric magnetic field extrapolations are therefore commonly used for estimations of the value and structure of the coronal magnetic field. Estimations show a low value of the ratio of gas pressure to magnetic pressure (plasma $\beta$) in the low and middle corona (Gary 2001). This means that the magnetic field governs plasma distributions and motions. Therefore, coronal structures depict the structure of the coronal magnetic field.

Unfortunately, we rarely see flux ropes clearly defined in the coronal structures, although some features could be interpreted as closely related to the flux ropes. Coronal cavities (Harvey 2001; Gibson *et al.* 2006, Reeves *et al.* 2012), that are visible in white-light and EUV coronal images at the base of coronal streamers and around prominences, seem to be the coronal structures that very likely correspond to flux ropes (Fig. 2). Depletion in white-light emission inevitably indicates plasma density lower within a cavity than in the surrounding corona. Sometimes this fact is considered as a hint on greater magnetic field strength within the cavity in accordance with the magnetohydrostatic pressure balance. However, in low-$\beta$ plasma this statement seems questionable.

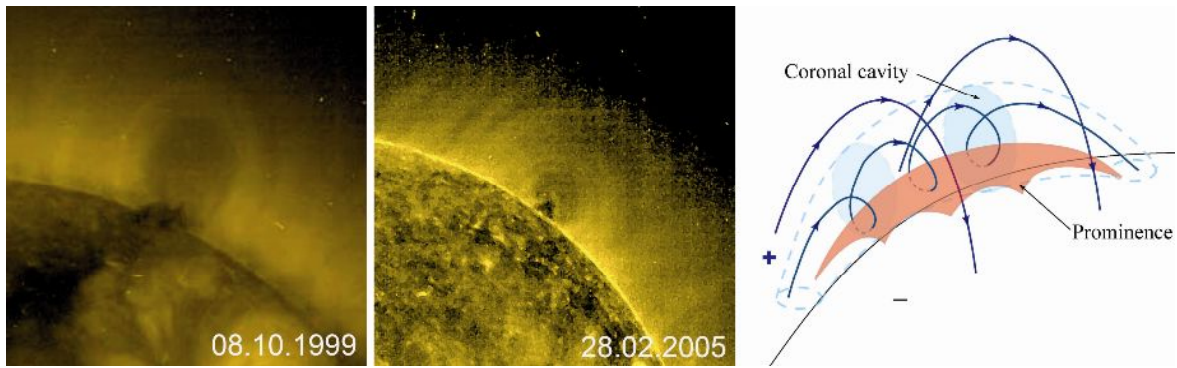

**Figure 2.** Coronal cavities observed in the SOHO EIT Fe XV 284 Å images and a scheme of a coronal flux rope. (Courtesy of SOHO/EIT Consortium.)

A coronal cavity is well recognized only when its axis is directed along the line-of-sight, otherwise it is screened by surrounding bright coronal loops. Many coronal cavities show a rather perfect round shape in their cross section. We can assume that the three-dimensional shape of the cavity is approximately a circular cylinder. Field lines of a flux rope have the form of helices. Their bottom parts serve as potential wells in the gravitational field where dense and cold plasma can be collected to form a prominence (or a filament, when observed on the disk). The prominence material can spread up to the center of the cavity until field lines reveal dips. Thus, solar prominences and filaments can be treated as tracers of flux ropes in the corona, and the top of a prominence can be considered as an indicator of the flux rope height above the photosphere. Régnier, Walsh, & Alexander (2011) reported observations with the Atmospheric Imager Assembly (AIA) onboard the Solar Dynamics Observatory (SDO) showing a polar crown cavity as a density depletion at the bottom of which the polar crown filament material lies indicating the existence of a magnetohydrostatic equilibrium. The filament material is drained down along the polar crown cavity by gravity and sustained by the action of the upward-directed magnetic field curvature force. The cold and hot coronal plasma are located at a similar location along the same field lines.

Since prominence material does not always fill the whole length of the helical flux tubes, the general helical geometry may not be fully revealed by the prominence or filament shape. Nevertheless, the twisted structure is visible in some of them, especially when they are activated or erupting (Fig. 3 and Fig. 4). During filament activation, plasma can spill over the upper parts of the flux tubes revealing the helical structure of the flux rope surrounding the filament. When viewed edge-on, along the flux rope axis, moving prominence material shows rotation around the axis (Fig. 5) like in terrestrial tornadoes (Wang & Stenborg 2010; Su *et al.* 2012; Li *et al.* 2012; Wedemeyer-Böhm *et al.* 2012). Sometimes helical threads of plasma filling an activated flux rope are visible on the disk (Gary & Moore, 2004; Kumar *et al.*, 2010; Joshi *et al*, 2014, see also Section 5) or close to the limb (Patsourakos, Vourlidas, & Stenborg 2013; Cheng *et al*. 2014a).

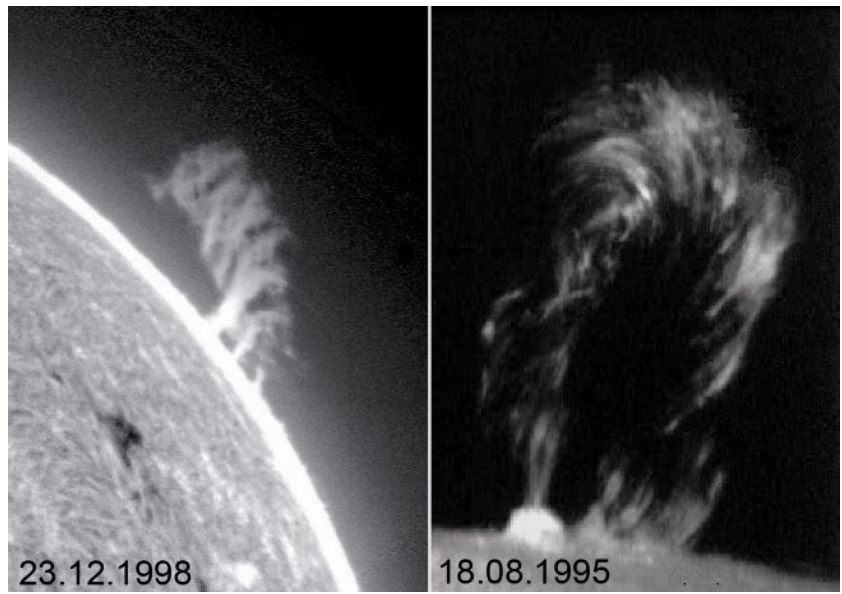

**Figure 3.** Twisted prominences. (Courtesy of Big Bear Solar Observatory.)

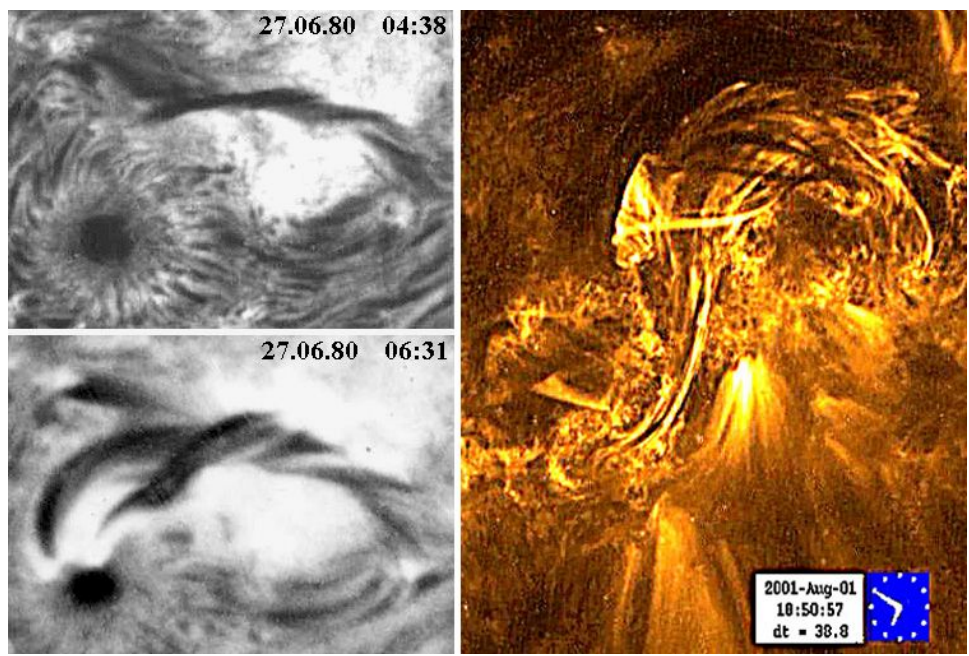

**Figure 4.** Twisted filaments. (Courtesy of Crimea Astrophysical Observatory and TRACE Consortium.)

Coronal sigmoids observed in soft X-rays (SXR) using the *Yohkoh* images and less definitely in EUV (Pevtsov, Canfield, & Zirin 1996; Aurass *et al*. 1999; Canfield, Hudson, & McKenzie 1999; Moore *et al*. 2001) also indicate the presence of high magnetic stresses and electric currents. Typically the central part of sigmoids is approximately aligned with a neutral line of the photospheric field. Filaments are present in most cases below sigmoidal coronal structures. Transient sigmoids become bright only shortly prior or during early, impulsive stages of flares. There is the opinion that transient sigmoids do not generally show the erupting flux ropes themselves but indicate the formation of the current sheet below them (Titov & Démoulin 1999; Kleim, Titov, & Török, 2004; Green *et al*. 2007). Observations are now considerably extended and improved our understanding and credit goes to new space missions like *Hinode*, STEREO and SDO. However, the basic rather large scale characteristics as described before do not change, although a lot of details could now be worked out with the need of considerable theoretical efforts before a clearer picture emerges.

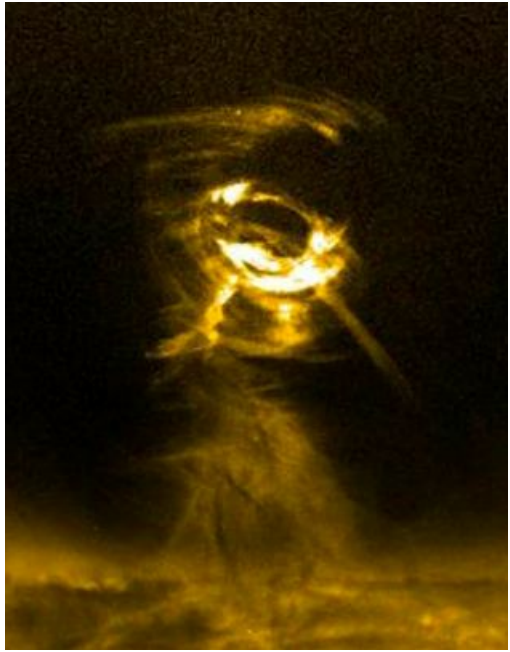

**Figure 5.** Tornado structure in the SDO/AIA 171Å channel on 25 September 2011 at 12:20 UT. (Courtesy of the AIA science team.)

### 3. Equilibrium and Stability of Flux Ropes in Coronal Magnetic Field

According to the magnetohydrodinamic equations, the equilibrium condition for a flux rope can be written (Izenberg, Forbes, & Démoulin 1993)

$$\frac{1}{c}\int_V (\mathbf{j}\times\mathbf{B}_f)\,dv + \frac{1}{c}\int_V (\mathbf{j}\times\mathbf{B}_e)\,dv - \oint_S p\,\mathbf{n}\,ds + M\mathbf{g} = 0 \;, \qquad (2)$$

where $V$ is the volume occupied by the flux rope, $S$ is the surface of the flux rope, $\mathbf{j}$ is the flux-rope current density, $\mathbf{B}_f$ and $\mathbf{B}_e$ are the magnetic fields due to the internal current of the flux rope and the external currents outside the flux rope, respectively, $M$ is the plasma mass containing within the flux rope. In many early 2D models, the cross-section of a flux rope was assumed to be small enough that the external field $\mathbf{B}_e$ was effectively uniform within the flux rope. With this assumption the equilibrium condition (2) can be decomposed into two separate conditions. One describes the global balance of forces per unit length acting on the flux rope as a whole

$$\frac{1}{c}\mathbf{I}\times\mathbf{B}_e + m\mathbf{g} = 0 \;, \qquad (3)$$

where $\mathbf{I}$ is the total electric current through the flux-rope cross section $A$ :

$$\mathbf{I} = \int_A \mathbf{j}\, ds \,.\tag{4}$$

The other equation describes the internal equilibrium within the flux rope:

$$\frac{1}{c}\mathbf{j} \times \mathbf{B}_f = \nabla p \,.\tag{5}$$

For a typical solar filament's electric current of $10^{11}$ A (Ballester 1984; Kulikova *et al.* 1986; Vrsnak *et al.* 1988; Srivastava, Ambastha, & Bhatnagar 1991) and coronal magnetic field of 10 G, a typical filament mass per unit length of $10^5$ g yields the gravitational force negligible in comparison with the electromagnetic force. The Lorentz force in Equation (3) vanishes when $\mathbf{B}_e = 0$. In equilibrium, a very thin flux rope (linear current) should be located at a null point of the external magnetic field. For the coronal magnetic field created by sub-photospheric currents the null point is expected to be of X-type or saddle-like (Fig. 5).

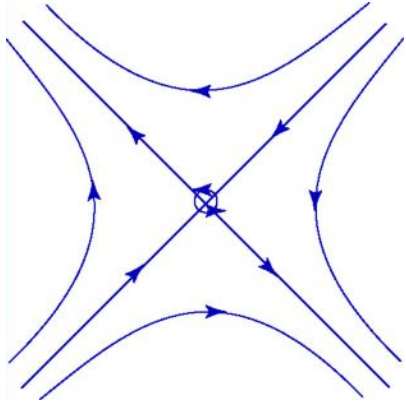

**Figure 5.** Magnetic field lines around an X-type or saddle-like null point with a linear electric current at the null.

A linear electric current at a null point is in equilibrium, but the equilibrium is not stable. For the shown geometry in Fig. 5, a small displacement in a vertical direction leads to the appearance of a restoring force, while a displacement in a horizontal direction leads to a net force pulling the current off the X-point. Stability can be achieved if one takes into account stabilizing action of the photospheric boundary. It acts like a rigid wall of a metallic vessel in a nuclear fusion installation. Kuperus & Raadu (1974) were first who paid attention to the action of dense photospheric plasma on the coronal current. van Tend & Kuperus (1978) showed first that there is a critical height for stable flux rope equilibria at which the background coronal magnetic field decreases faster than the inverse height.

In the simplest model with the flux rope considered as a straight linear current, the vertical component of Equation (3) is (van Tend & Kuperus 1978; Molodenskii & Filippov 1987)

$$F = \frac{I^2}{c^2 h} - \frac{I}{c} B_e(h) - mg = 0 \,,\tag{6}$$

where $h$ is the height of the electric current above the photosphere. The first term represents the repulsion between the coronal current and the mirror current. The second term represents the Lorentz force, with which magnetic field of sub-photospheric sources attracts the coronal current to the photosphere. Since the mirror current does not create a horizontal force acting on the coronal current, horizontal equilibrium is achieved at a point where the magnetic field lines are horizontal, i.e. above a polarity inversion line. Neglecting the weight of the filament, the vertical balance is reached at a height $h_0$ determined by the equation

$$\frac{I}{ch_0} = B_e(h_0). \tag{7}$$

The equilibrium height of the current is related to the current strength. The stronger the electric current, the higher the equilibrium position. The vertical stability is determined by the sign of the second derivative of the potential energy, which should be positive for stability,

$$\left.\frac{d^2W}{dh^2}\right|_{h_0} = -\left.\frac{dF}{dh}\right|_{h_0} > 0 \tag{8}$$

The horizontal stability in this model requires the curvature of the background field lines to be directed downwards. If we assume that the background magnetic field changes with height within some interval as a power-law function

$$B_e(h) = Ch^{-n}, \tag{9}$$

the degree $n$, called also the "decay index", should be less than unit (Filippov & Den 2000, 2001).

The height $h_c$ where $n = 1$ is the limit for a stable equilibrium. The critical height characterises the scale of the background magnetic field. We can calculate this parameter, for example, in the current-free approximation using photospheric magnetic field measurements by solving the equation

$$h_c \left.\frac{dB_e}{dh}\right|_{h_c} = -B_e. \tag{10}$$

Of course, the magnetic field within a flux rope is not potential but for the analyses of the equilibrium of the coronal current as a whole, we just have to use the external magnetic field generated by sub-photospheric currents, excluding the field of the coronal current and the field of the mirror current.

When an electric current is curved, an extra force is present, called the hoop force (e.g. Bateman 1978). Equilibrium of a toroidal pinch was studied by Shafranov (1966). The equations of the force balance along the major torus radius $R$ and along the minor radius $a$ are

$$\frac{B_a^2}{4\pi}\left\{\ln\frac{8R}{a} - 1 + \frac{l_i}{2}\right\} + \bar{p} - p_e + \frac{B_e^2 - \bar{B}_i^2}{8\pi} + \frac{B_z B_a R}{2\pi a} = 0, \tag{11}$$

$$\bar{p} - p_e + \frac{B_e^2 - \bar{B}_i^2}{8\pi} - \frac{B_a^2}{8\pi} = 0, \tag{12}$$

where $B_a$ is the poloidal magnetic field, $\overline{B}_i$ is the averaged internal toroidal field, $B_z$ is the external field perpendicular to the plane of the torus' axis, $\overline{p}$ is the averaged internal plasma pressure, $p_e$ is the external plasma pressure, $l_i$ is the internal inductance. Shafranov (1966) stressed the need of the $B_z$ component of the external magnetic field for equilibrium of the toroidal pinch. The Lorentz force of the toroidal plasma current in the field $B_z$ (the last term in Equation (11)) provides the inward force that holds the plasma torus in equilibrium. A number of eruptive event models were based on preexisting flux ropes with toroidal symmetry in ejection source regions (Lin *et al.* 1998; Chen 1989; Titov & Démoulin 1999; Roussev *et al.* 2003; Kliem & Török 2006; Isenberg & Forbes 2007; Fan & Gibson 2007; Olmedo & Zhang 2010).

The stability of the Shafranov equilibrium has been considered by Bateman (1978), who found that the ring current is unstable against expansion if the external field decreases sufficiently rapidly in the direction of the major torus radius $R$. Kliem & Török (2006) called the related instability as "torus instability" and showed, following Bateman (1978), that it occurs when the background magnetic field decreases along the expanding flux ropes' major radius $R$ faster than $R^{-1.5}$. They analyzed cases where the electric current $I$ was held constant or fixed by the conservation of the total magnetic flux within the torus hole.

As the photospheric magnetic field distribution and the corresponding coronal field gradually change, a coronal flux rope evolves quasi-statically along a sequence of stable equilibrium unless it encounters a critical point on the equilibrium manifold. The catastrophe then occurs by a loss of equilibrium (Priest & Forbes 1990; Forbes & Isenberg 1991; Isenberg, Forbes, & Démoulin 1993; Forbes & Priest 1995; Lin *et al.* 1998; Lin & Forbes 2000; Lin & van Ballegooijen 2002; Schmieder, Démoulin, & Aulanier, 2013; Longcope & Forbes 2014). The transition of an equilibrium flux rope to a state of non-equilibrium can be treated as instability when the evolution of a small perturbation acting on an equilibrium at any point on the equilibrium manifold is considered. Démoulin & Aulanier (2010) showed that the loss of equilibrium and the torus instability are two different views of the same physical mechanism. They identified that the same physics was involved in the instability of circular and straight current channels. Kliem *et al.* (2014), using a toroidal flux rope in an external bipolar or quadrupolar field as a model for the current-carrying flux, verified that the catastrophe and the torus instability occur at the same point. Thus, they are equivalent descriptions for the onset condition of solar eruptions.

We see that the stable equilibrium of a flux rope is possible only if the background field does not decrease with height rapidly, or the decay index of the ambient magnetic field (Filippov & Den 2000, 2001)

$$n = -\frac{\partial \ln B_e}{\partial \ln h}, \qquad (13)$$

does not exceed a critical value $n_c$. The exact value of $n_c$ depends on parameters of a flux rope model. For a thin straight current channel $n_c = 1$, while for a thin circular current channel $n_c = 1.5$. The difference decreases if current-channels are rather thick. Démoulin & Aulanier (2010) showed that a critical decay index $n_c$ has similar values for both the circular and straight current channels in the range 1.1 - 1.3, if a current channel expands during an upward perturbation, and in the range 1.2 - 1.5, if a current channel would not expand.

Filippov & Den (2000, 2001), Filippov & Zagnetko (2008) calculated on the basis of photospheric magnetograms the decay index in the vicinity of filaments, using the potential magnetic field approximation. They defined a critical height $h_c$ as the height where $n = 1$ or, in other words, used $n_c$ for a straight current channel. They compared the measured heights of stable and eruptive filaments with the critical heights and found that the heights of stable filaments are usually well below the critical heights, while the heights of filaments just before their eruption are close to the instability threshold (Fig. 6).

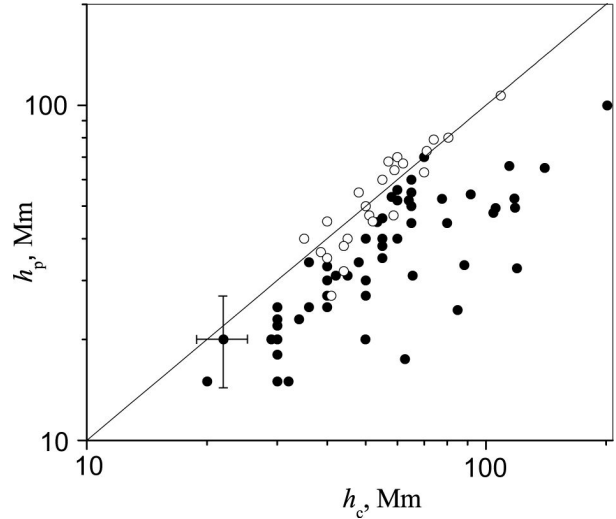

**Figure 6.** Observed filament heights above the limb $h_p$ versus the critical heights of stable filament equilibrium $h_c$. The solid circles correspond to the filaments which safely passed the west limb. The open circles correspond to the filaments which disappeared from the disk before they reach west limb. The straight line corresponding to equality of these quantities is the stability boundary.

The decay index was computed in regions above the photosphere from a potential field extrapolation (Liu 2008; Guo *et al.* 2010; Kumar *et al.* 2012; Nindos, Patsourakos, & Wiegelmann 2012; Xu *et al.* 2012) or a nonlinear force-free field extrapolation (Liu *et al.* 2010; Cheng *et al.* 2011; Savcheva, van Ballegooijen, & DeLuca 2012) in order to find the difference in coronal magnetic fields for failed eruptions and full eruptions, a threshold of flux-rope instability, the relationship between the decay index and CME speed, and so on. The value of the decay index above 1.5 near a filament was considered as the manifestation of the necessary condition for the torus instability (Kumar *et al.* 2012; Zuccarello *et al.* 2013, 2014). However, more accurate estimations show the value of the critical decay index in some of these events to be close to unity (Filippov 2013; Filippov *et al.* 2014).

## 4. Chromospheric Reflection of Coronal Equilibrium Conditions

Despite the progress in 3D numerical simulations, simple 2D models are still widely used in general to understand of fundamental properties of flux-rope equilibrium and stability. In qualitative, schematic considerations, different initial equilibrium configurations are presented. In particular, there are many cartoons (Pneuman 1983; Malherbe & Priest 1983; Anzer & Priest 1985; Priest 1990; Priest & Forbes 1990) showing an inverse polarity filament with an X-type singular point below it (Fig. 7(b)). This figure-of-eight-type configuration is sometimes referred as the Kuperus–Raadu model. However, in the Kuperus and Raadu (1974) original paper only the configuration shown in Fig. 1(a) was presented. This is also called the bald-patch separatrix surface configuration (BPS), while the former is called the hyperbolic flux tube configuration (HFT) (Titov, Priest, & Démoulin 1993; Titov, Hornig, & Démoulin 2002; Kliem, Török, & Forbes 2011). The observation that the coronal mass ejection acceleration phase usually coincides with the soft X-ray flare rise phase, which was first demonstrated by Zhang *et al.* (2001), sometimes is interpreted as evidence of the HFT configuration in the corona prior to the eruption (Kliem, Török, & Forbes 2011). Field-line reconnection, usually associated with a flare, can start at the X-type structure immediately after the flux-rope instability begins.

A null point can appear in the corona due to the complicated photospheric magnetic-field distribution (at least a quadrupolar structure). The two configurations shown in Fig. 7 contain only a bipolar region and a coronal current. The null point exists in the corona because the fields of the bipolar arcade and that of a coronal current are superposed. The detailed analysis leads to conclusion that the configuration shown in Fig. 7(b) is not in stable equilibrium and cannot be used as a pre-eruptive state for solar eruptive events.

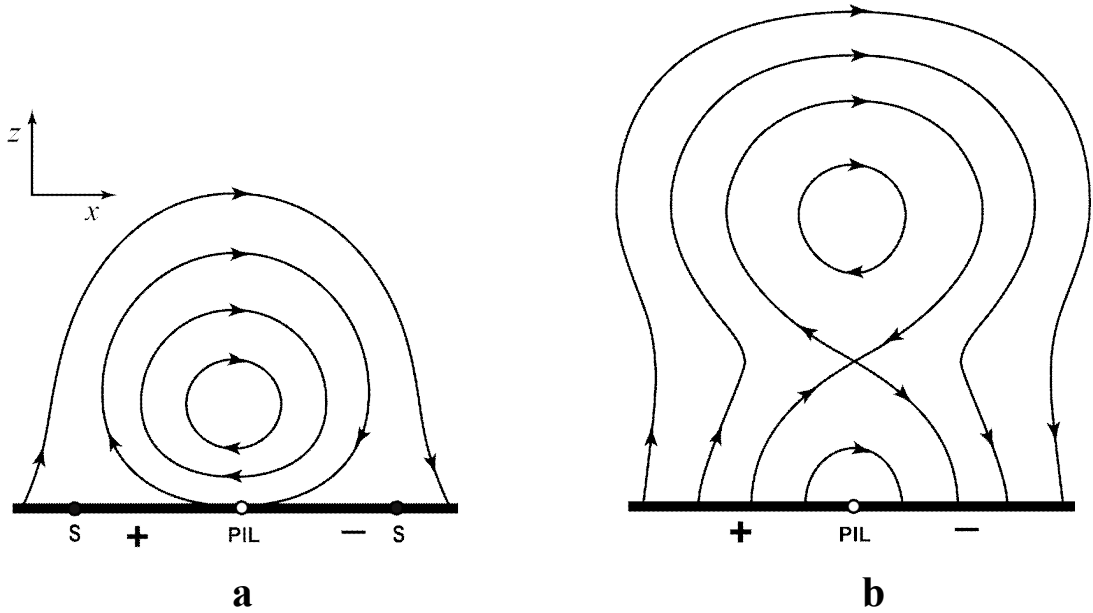

**Figure 7.** Basic magnetic topology for the van Tend and Kuperus (1978) model of prominence support (a) and the configuration with an X-point between a flux rope and the photosphere (b).

In principle, equilibrium conditions in the corona are reflected in the distribution of the photospheric magnetic field. It is known that the MHD equation of momentum conservation can be written as (Landau, Lifshits, & Pitaevsky 1984; Kuperus & Raadu 1974)

$$\frac{\partial \rho v_i}{\partial t} = -\frac{\partial \Pi_{ik}}{\partial x_k}, \qquad (14)$$

where $x_i = (x, y, z)$ and summation over repeated indices is to be understood. $\Pi_{ik}$ is a symmetric tensor of second rank with the components

$$\Pi_{ik} = \rho v_i v_k + p \delta_{ik} - \frac{1}{4\pi}\left(B_i B_k - \frac{1}{2} B^2 \delta_{ik}\right), \qquad (15)$$

In equilibrium, **v** = 0 and in a low-$\beta$ coronal plasma, we can neglect the gas-pressure term in Equation (15). Integrating Equation (14) with the zero left-hand side over the volume bounded by the surface $z = 0$ of the chromosphere (assumed here as a flat surface) and a hemisphere of radius $R$, we obtain after reducing the volume integral to a surface integral (Molodenskii & Filippov 1989)

$$\oint \left( B_i B_k - \frac{1}{2} B^2 \delta_{ik} \right) \mathrm{d}s_i = 0. \tag{16}$$

Since for a concentrated source of the magnetic field, $B$ falls off at large distances as $R^{-3}$, we may integrate only over the surface $z = 0$ of the chromosphere. The equilibrium conditions along $x$-, $y$-, and $z$-axes for the chromospheric magnetic-field components are then given by

$$\int_{-\infty}^{\infty}\int_{-\infty}^{\infty} B_y B_z \,\mathrm{d}x\mathrm{d}y = 0, \tag{17}$$

$$\int_{-\infty}^{\infty}\int_{-\infty}^{\infty} B_x B_z \,\mathrm{d}x\mathrm{d}y = 0, \tag{18}$$

$$\int_{-\infty}^{\infty}\int_{-\infty}^{\infty} \left( B_z^2 - B_x^2 - B_y^2 \right) \mathrm{d}x\mathrm{d}y = 0. \tag{19}$$

These conditions for photospheric force-free fields were first obtained by Molodensky (1974). He also showed that these equations are satisfied in sunspots within the accuracy of measurements. Equation (19) describes the vertical equilibrium; this means that the mean-square value of the vertical field should be equal to the mean-square value of the horizontal field.

Let us represent field **B** as a sum of field $\mathbf{B}_0$ of sub-photospheric sources and field **b** of coronal currents

$$\mathbf{B} = \mathbf{B}_0 + \mathbf{b}. \tag{20}$$

Field $\mathbf{B}_0$ is the potential field and satisfies Equations (17) – (19). Due to the photospheric diamagnetism

$$b_z(x,y,0) = 0. \tag{21}$$

Substituting Equation (20) into (19) and using the boundary condition (21) and potentiality of $\mathbf{B}_0$, we obtain (Molodenskii & Filippov 1989)

$$\int_{-\infty}^{\infty}\int_{-\infty}^{\infty} \left[ b_x(2B_{0x} + b_x) + b_y(2B_{0y} + b_y) \right] \mathrm{d}x\mathrm{d}y = 0. \tag{22}$$

For the two-dimensional topology presented in Fig. 7 we may set

$$b_y = 0 \tag{23}$$

and to have the equilibrium condition in the form

$$\int_{-\infty}^{\infty} b_x (2B_{0x} + b_x)\, \mathrm{d}x = 0, \tag{24}$$

or

$$\int_{-\infty}^{\infty} b_x (B_{0x} + B_x)\, \mathrm{d}x = 0. \tag{25}$$

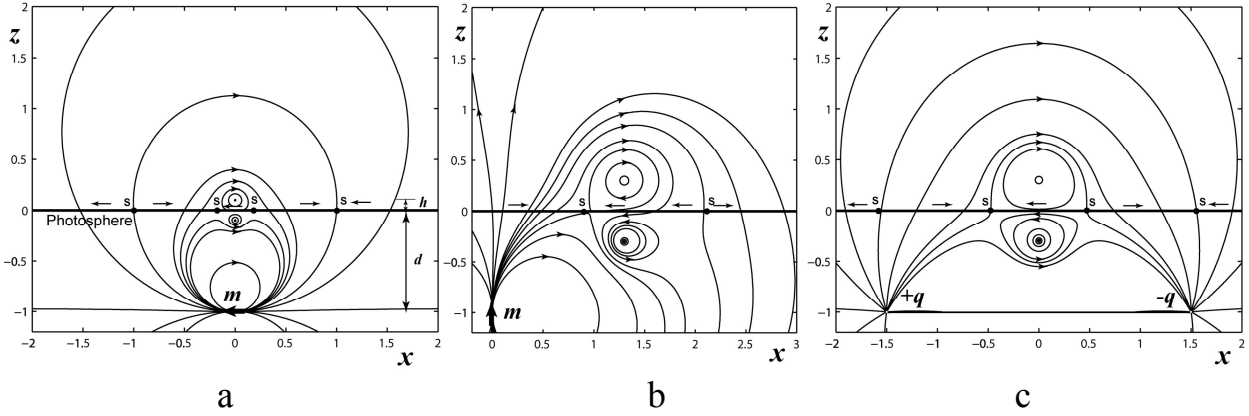

**Figure 8.** Field lines of simple 2D models of line-current equilibrium in the field of a horizontal dipole (a), a vertical dipole (b), and two separated charges (c). Small arrows show the direction of the photospheric horizontal field. S denotes the separatrices.

For a single flux-rope in the corona, as in Fig. 7, the integration should be performed within limits of the order of the scale of field $b$, which is about the height $h$ of current above the chromosphere. The component $b_x$ has the same sign over the whole chromosphere. The component $B_{0x}$ can change the sign, however, if the sub-photospheric currents are located rather deep below the surface at a depth $d$ (see Fig. 8(a)), the lines of the $B_{0x}$ sign change are located at a distance on the order of $d$ away from a photospheric polarity inversion line. When $h \ll d$, the sign of $B_{0x}$ is also constant within the integration region $-h < x < h$. Then, as seen from Equation (24), the signs of $B_{0x}$ and $b_x$ should be opposite, which is inherent for inverse polarity filaments (Leroy 1989; Paletou & Aulanier 2003). Equation (25) needs the total field $B_x$ to change sign at least in some part of the integration region. This means the existence of separatrices S in the horizontal field distribution (Fig. 7(a)). Just below the coronal current position, the direction of the entire field $B_x$ is opposite to that of the sub-photospheric sources $B_{0x}$. Clearly, the configuration presented in Fig. 7(b) does not meet this condition.

If $h \gg d$, the situation is not as clear as was in previous case. In this case, $B_{0x}$ changes sign within the integration region, therefore $B_x$ could in principle be unidirectional below the flux rope. However, in this condition, the flux rope equilibrium cannot be stable (Molodenskii & Filippov 1987; Priest & Forbes 1990).

From the general relationship it is the necessity for flux-rope field near the surface of the chromosphere to be opposite to the field of sub-photospheric sources. For wide variety of photospheric fields, the total horizontal field in the chromosphere changes sense at some lines, separatrices, parallel to the polarity inversion lines. Filippov (2013) considered three simple models of flux-rope equilibrium shown in Fig. 8 and found that except some special conditions the separatrices are present at the chromospheric level at both sides of a polarity inversion line. Possibly the most relevant to typical solar conditions is the background magnetic field represented by two "charges" $\pm q$ (Fig. 8(c)). If the depth $d$ in the model is much less than the distance $2a$ between the charges, there is a pair of separatrices above the charges with coordinates $x_s = \pm a$ and a pair of $h$-dependent separatrices :

$$x_s = \pm h\sqrt{\frac{3a^2 - h^2}{5h^2 + a^2}} . \qquad (26)$$

The separatrices can be recognized in many filtergrams near the solar filaments. Chromospheric fibrils show, with 180º ambiguity, the direction of the horizontal magnetic-field component in the chromosphere (Foukal 1971; Zirin 1972). Near the filaments, they reflect the specific magnetic configuration that supports filament material high in the corona. A fibril pattern below a filament is so peculiar that it received a special name, "a filament channel" (Martres, Michard, & Soru-Escout 1966; Gaizauskas *et al*. 1997; Martin 1998). Fibrils in the channel run nearly parallel to the polarity boundary, manifesting the presence of the strong-field component along the boundary. Filament barbs, which represent some prominent threads forming the filament's body, are parallel to the chromospheric fibrils just below the filament when viewed from above (Martin 1998). Using information about magnetic-polarity sense from a magnetogram and the field continuity, one can derive the direction of the horizontal magnetic field in a filtergram. Figure 9 clearly shows two lines that separate areas with opposite direction of a component perpendicular to the filament axis. Since a component parallel to the filament axis dominates in the filament channel, fibrils near the separatrices are aligned with the filament. They form a conspicuous "herring-bone structure" (Filippov 1994, 1995). From this feature, separatrices near filaments can be easily recognized.

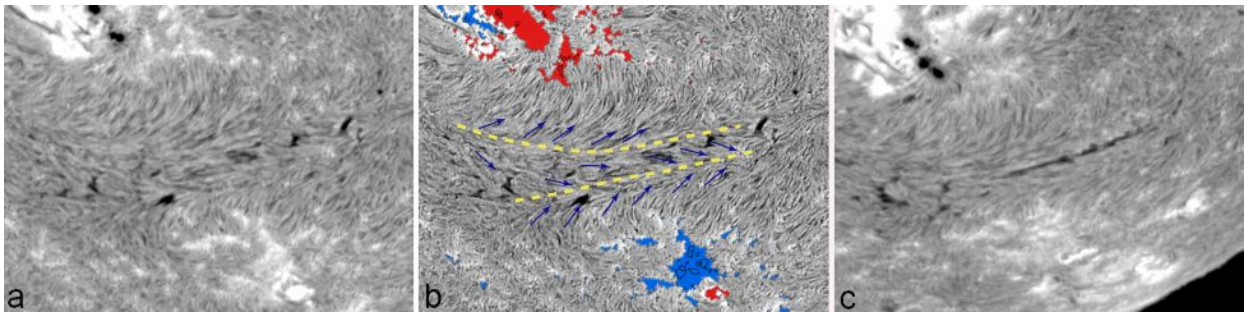

**Figure 9.** (a) BBSO Hα filtergram on 10 November 2001. (b) The same filtergram with an unsharp–mask filter applied and with the photospheric magnetic-field concentrations stronger than ±50 gauss (G) from a SOHO/MDI magnetogram superposed. Dashed yellow lines represent separatrices, dark blue arrows show the orientation of the horizontal magnetic field component deduced from the fibril pattern. Red areas represent negative polarity, while blue areas represent positive polarity. (c) BBSO Hα filtergram on 12 November 2001. The filtergrams are rotated by 45º clockwise to make the filament axis to nearly horizontal in the frames. The images are 620" × 470" across. (Courtesy of Big Bear Solar Observatory and SOHO/MDI consortium).

The filament in Fig. 9(a) is very faint. Only a chain of small fragments is visible along the filament spine. However, owing to the filament transparency, the fibril pattern below it is more sharply defined. Nevertheless, the filament was visible every day during its passage across the disk because of the solar rotation. In the days after 10 November, the filament became more solid although it was narrow and low.

The photospheric sources of the magnetic field in Fig. 9 are concentrated and located rather far from the filament channel. The configuration is more similar to the model shown in Fig. 8(c). The distance between the separatrices according to Equation (26) is about $2x_s = 2h\sqrt{3}$. In Fig. 9(b), the distance between the separatrices is about 29 Mm. The height of the filament can be estimated in the filtergram taken on 14 November, when the filament was close to the limb. It is comparable with the width of the filament in Fig. 9(c) and reaches nearly 9 Mm. Therefore, the foregoing relationship holds with good accuracy.

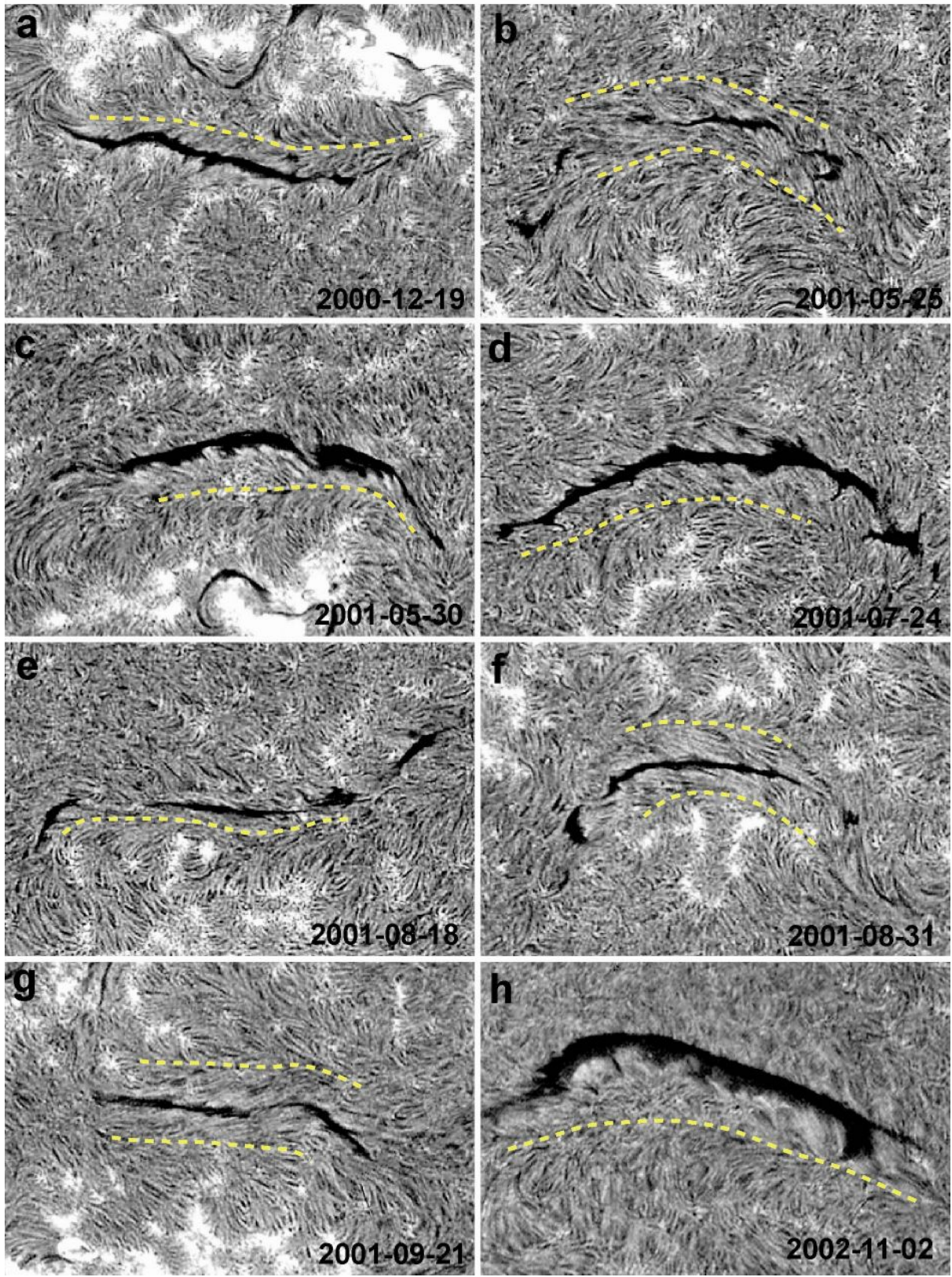

**Figure 10.** BBSO Hα filtergrams showing separatrices in the fibril distribution near filaments. The size of each frame is 410" × 270". The filtergrams are rotated at different angles to make the filament axis nearly horizontal in frames. An unsharp mask filter was applied to see the structure more clearly. (Courtesy of Big Bear Solar Observatory).

Separatrices near filaments can be recognized to a greater or lesser extent in many filtergrams (see Fig. 10). Of course, the clearness of the structure depends on the regularity of the surrounding magnetic field. In some filtergrams in Fig. 10, only one separatrix is clearly discernible. It could be related to a highly non-symmetric magnetic-field distribution relative to a polarity inversion line, a strong field on the one side and a weaker field on the other side, as in the case of the vertical dipole (Fig. 8(b)), or a too strong vertical-field component at the expected separatrix location that prevents one from recognizing the horizontal field structure in a fibril pattern.

## 5. Flux-Rope Helicity and Filament Chirality

The axial component of the filament magnetic field defines two classes of filaments, depending on the direction of the axial component: a filament is called dextral if this component is directed towards the right when the filament is viewed from the side of the positive background polarity, and sinistral if the direction of the axial component is opposite to this (Martin, Billimoria, & Tracadas 1994). Analysis of the fine structure of filaments shows that the thin threads are rotated through a small angle clockwise to the axis in dextral and counterclockwise in sinistral filaments. This makes it possible to determine the class of a filament (the filament chirality) from its visual appearance, without information on the magnetic fields (Pevtsov, Balasubramaniam, & Rogers 2003). Dextral filaments are predominantly located in the northern hemisphere and sinistral filaments in the southern one, and this dominant location does not vary from cycle to cycle (Martin, Billimoria, & Tracadas 1994; Zirker *et al.* 1997).

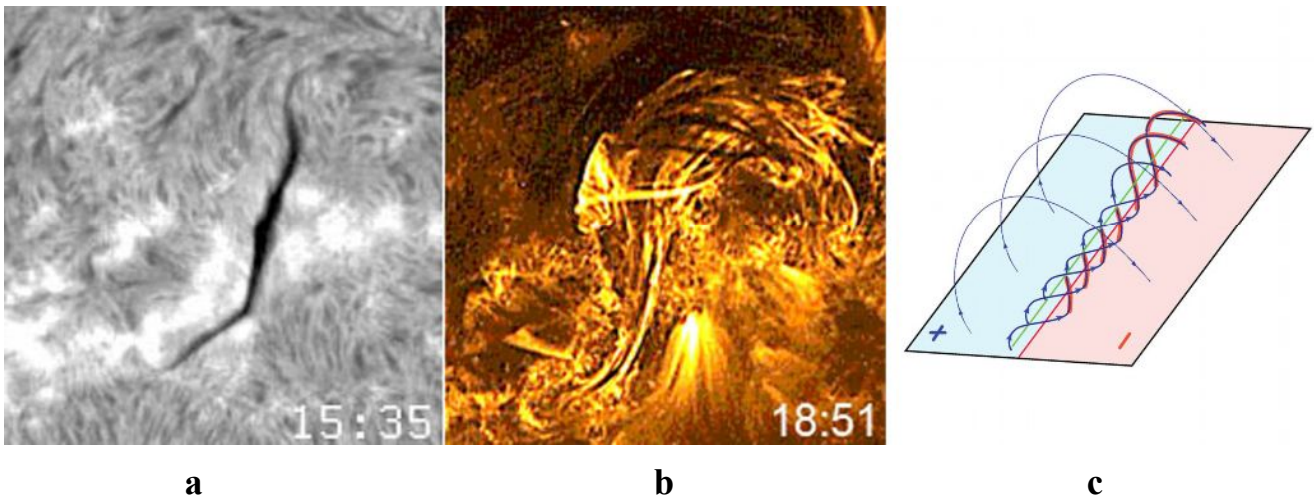

        **a**                                **b**                            **c**

**Figure 11.** Activation of a sinistral filament observed on August 1, 2001. (a) Hα filtergram at 15:35 UT (Big Bear Solar Observatory), (b) TRACE UV image in the 171 ˚A channel, (c) schematic of the sinistral filament located inside a magnetic-flux rope. The filament plasma fills the lower volumes of the cylindrical magnetic tubes (thick red lines). During the filament activation, the material can flow through upper volumes, as is shown in the upper part of the plot. (Courtesy of Big Bear Solar Observatory and TRACE team. TRACE is a mission of the Stanford-Lockheed Institute for Space Research (a joint program of the Lockheed-Martin Advanced Technology Center's Solar and Astrophysics Laboratory and Stanford's Solar Observatories Group) and part of the NASA Small Explorer program.).

On the other hand, the orientation of the fine structures of filaments and of chromospheric fibrils located beneath the filaments correspond to the direction of the magnetic component, which is transverse to the neutral line and is opposite to that determined from the photospheric polarities (Filippov 1998). The measurements of magnetic fields in prominences also reveal a predominantly inverse polarity. The source of both the inverse and axial fields must be located near the neutral line, since the fibril orientation more or less matches the direction of the potential field at comparatively short distances from this line, namely, beyond the filament channel.

Most of the structural features of filaments are consistent with the configuration of the surrounding magnetic field being in the form of magnetic flux ropes. The lower portions of helices form the fine structure of the filaments observed in chromospheric lines. During a filament activation or eruption, the upper portions of the helical magnetic tubes may contain hot plasma with larger scale heights and moving cold plasma that is far from hydrostatic equilibrium (upper part of Fig. 11(c)). The eruptive prominences demonstrate the helical morphology most clearly. Figure 11(a) presents the filtergram of a sinistral filament. The fine structural elements deviate from the filament axis counterclockwise. Figure 11(b) shows the image of the same filament obtained three hours later, during the activation. We can clearly see the threads overlapping underlying elements and being deflected clockwise from the axis. Thus, this filament demonstrates magnetic structure corresponding to a right-handed cylindrical helix. After the activation, which persisted for two hours and demonstrated intense internal motions with material flowing through the upper portions of the helix, the filament came back to a quiet state and was almost restored to its previous form.

The direction of twisting of the helix corresponds to the direction of the electric current assumed in models of inverse polarity filaments. The ambient field generated by the dominant photospheric sources applies to this current a force directed toward the photospheric polarity inversion line. For a stable equilibrium in the corona, we must take into account the diamagnetism of the dense photosphere.

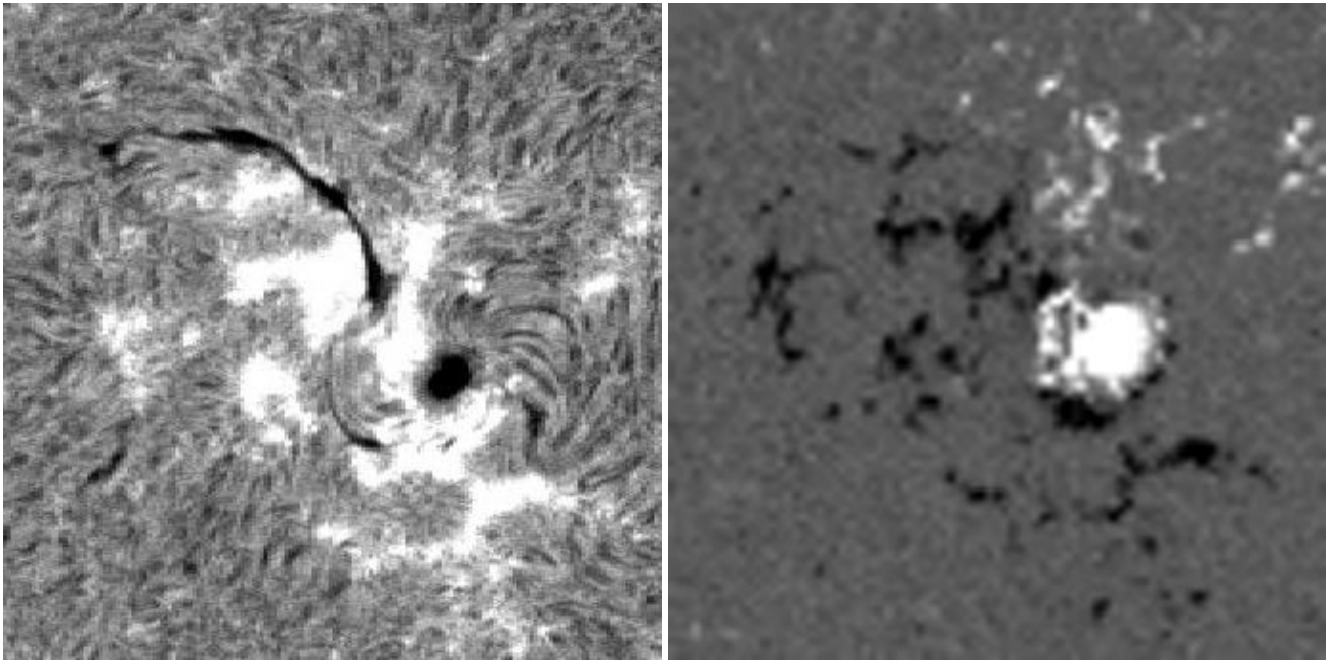

**Figure 12.** Hα filtergram of a dextral filament and sunspot with a vortical superpenumbra (at the left) obtained on June 17, 1998 at 18:43 UT (Big Bear Solar Observatory) and the magnetogram of this region obtained at 14:24 UT (SOHO/MDI). (Courtesy of Big Bear Solar Observatory and SOHO/MDI Consortium).

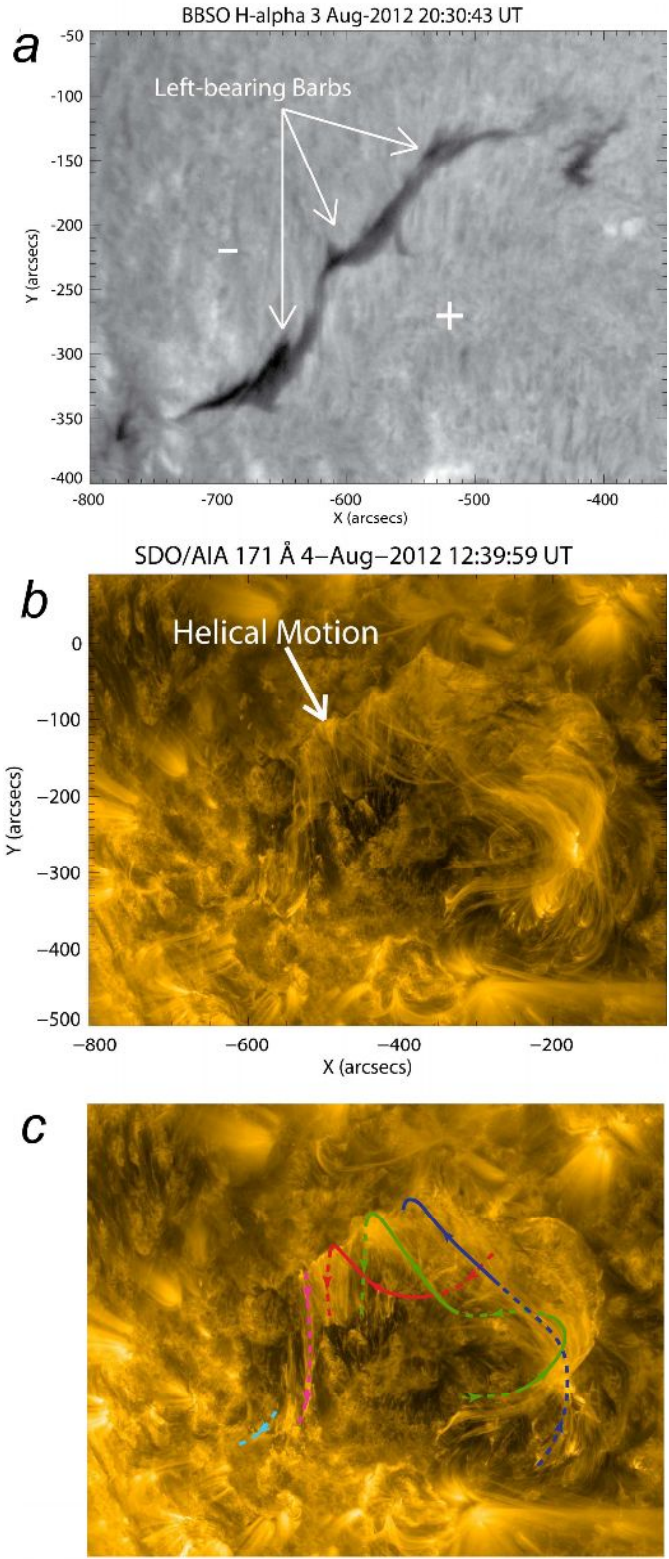

**Figure 13.** (a) Hα image of the filament on 2012 August 3 at 20:30:43 UT (Big Bear Solar Observatory). The + and −signs indicate the positive and negative polarity regions. White arrows indicate the left-bearing barbs of the filament. (b) SDO/AIA 171Å image at 12:39:59 UT showing the activation of the right-handed helically twisted flux rope containing the filament. (c) SDO/AIA 171Å image at 12:14:35 UT with drawn sections of flux rope field lines that can be followed from the visible threads. Field-line portions that lie above the flux rope axis are shown as solid lines, while those that lie below the flux rope axis are shown as dashed lines. (Courtesy of Big Bear Solar Observatory and of the AIA science team.).

The sign of the helicity assumed for the filaments associated with sunspot superpenumbrae (Fig. 12) corresponds to the vorticity of the superpenumbra. According to the estimates of Kulikova *et al.* (1986, 1989), both the direction and magnitude of the electric currents in the sunspot and adjacent filament correspond to each other, so that the filament current can close at the photosphere through the sunspot. Figure 12 shows a sunspot of positive polarity that demonstrates a clear helicity of the superpenumbra twisted counterclockwise, which is conjugate with a dextral filament. The filament chirality can be independently determined from both its fine structure and the surrounding photospheric polarities, taking into account the conjugacy of one of the filament ends with the sunspot of positive polarity. Rust & Martin (1994) noted the essentially unambiguous correspondence between the direction of the sunspot vortices and the chirality of the associated filaments. In addition, both counterclockwise vortices and dextral filaments dominate in the northern hemisphere, while clockwise vortices and sinistral filaments dominate in the southern hemisphere.

A clear example of one-to-one correspondence between the filament chirality and the enveloping flux rope helicity was found by Joshi *et al.* (2014). Figure 13 shows the high-resolution image of the filament cold plasma observed in Hα line at Big Bear Solar Observatory on 3 August 2012. The filament has three barbs, which are marked by white arrows in Figure 13. In most cases, the barbs of a dextral (sinistral) filament are observed to be right (left) bearing (Martin 1998). The direction of thin threads within the main body of the filament often shows the direction of the magnetic field within the filament, which is related to chirality. From Figure 13, it is clear that the filament barbs are left bearing, which corresponds to sinistral chirality. Thin threads within the main body of the filament deviate counterclockwise from the filament axis, which also corresponds to sinistral chirality.

An eruption of a short eastern section of the huge filament caused the simultaneous activation of a large filament to the west of the eruption place and a large-scale flux rope containing the filament. The coronal filters of SDO/AIA 171Å provide the observation of the flux rope after the tracking of hot plasma along its various flux tubes (Fig. 13 (b), (c)). Some threads elongate showing the motion along the field lines, while others move laterally as a whole showing evolution of the flux rope magnetic field during the activation. Careful inspection shows that these features move below the main filament body and therefore below the flux rope axis. Subsequently, it becomes obvious that plasma rises up on the northern side of the flux rope and then moves to the southwest above the flux rope axis. As a result, field-aligned plasma motion demonstrates a clockwise rotation when viewed along the material moving away and, therefore, a right-handed helix. The rotation motion is clearly visible in the limb view along the axis of the flux rope. The EUVI instrument Sun Earth Connection Coronal and Heliospheric Investigation (SECCHI) on-board Solar Terrestrial Relation Observatory Behind (STEREO B) provides images in this projection (Fig. 14). The clockwise rotation of bright features within the bright border of the nearly circular section of the flux rope (pointed by the white arrow in Fig. 14) exactly corresponds to the field aligned motion observed on the disk. This helical plasma dynamics was recently modelled by stringent 3-D numerical simulation where near photospheric perturbations in the azimuthal component of the magnetic field generated the plasma as well as field dynamics that showed similar feature evolution (Murawski *et al.* 2014).

The observed rotation and the sign of helicity (positive) of the flux rope containing the sinistral filament strongly support the idea of dextral (sinistral) filaments being associated with flux ropes with negative (positive) helicity (Chae 2000; Rust 1999).

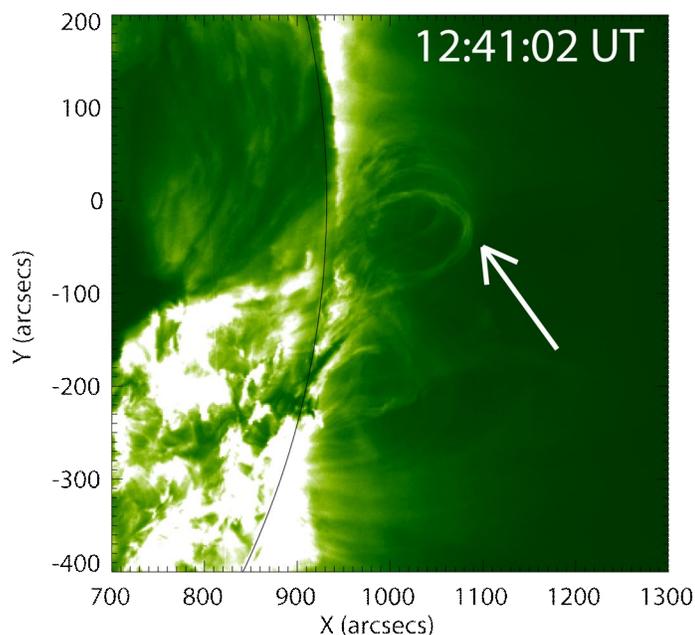
**Figure 14.** Nearly circular section of the flux rope observed by the STEREO/SECCHI/EUVI in 195Å channel.

## CONCLUSIONS

The space weather variability is mostly determined by the active processes on the Sun which disturb the interplanetary medium. Investigations of the last decades put coronal mass ejections on the first place among solar geo-effective phenomena. As a rule, the major geomagnetic substorms happen after fast coronal mass ejections (Kahler 1992; Gosling 1993), while interplanetary shocks generated by them accelerate charged particles to high energy producing radiation hazard (Kahler 2003). The most energetic events on the Sun are accompanied by both CMEs and solar flares. It is very likely that the two phenomena are interrelated.

Many observational facts and the theoretical analysis show that magnetic flux ropes are the most probable candidates for source regions of eruptive phenomena. They seem very promising structures to be able to store free magnetic energy in the corona. Although it is not so easy to recognize the flux ropes in the coronal structure, prominences and filaments associated with them are clearly seen in chromospheric and also coronal images. They are the best tracers of the flux ropes in the corona long before the beginning of an eruption. Sudden catastrophic loss of the flux rope equilibrium is the cause of filament eruptions and coronal mass ejections. The stability of the flux rope depends on the strength of the total electric current flowing within it and properties of the surrounding coronal magnetic field. The change of equilibrium conditions within the flux rope may also be accompanied with the flux emergence (a classical trigger of eruption; e.g, Kumar *et al*. 2011). There is a critical height for the stable flux-rope equilibrium in any given magnetic field. This parameter can be calculated on the basis photospheric magnetic field measurements. The comparison of the measured heights of prominences with calculated for the ambient magnetic fields critical heights can be the basis for predicting filament eruptions and the following CMEs.

In conclusion, the inherent plasma and magnetic field changes and the similar changes in the ambient corona may lead the dynamics and instability of the solar magnetic flux ropes. In spite of that, some canonical processes may also subject in making the magnetic flux ropes unstable (Török & Kleim 2005; Srivastava *et al*. 2010; 2013; Cheng *et al*. 2014b; Korsós, Baranyi, & Ludmány 2014; Korsós *et*

*al.* 2015, and references cited there). Better understanding the physics of solar magnetic flux ropes: their formation, identification, stability, and instability conditions are of great importance in solar and heliospheric physics as they are the precursor of huge coronal mass ejections and geomagnetic storms. The space-borne and ground observatories and major part of solar physics community are involved in understanding the space weather consequences, therefore, concentrate also on the study of the nature of large-scale solar magnetic flux-ropes, which will be very important in predicting the onsets of solar flares, eruptions, related dynamical processes, and geomagnetic storms.

## ACKNOWLEDGMENTS

This work was supported in part by the Russian Foundation for Basic Research (grant 14-02-92690) and in part by the Department of Science and Technology, Ministry of Science and Technology of India (grant INT/RFBR/P-165)